\documentclass[12pt]{revtex4-1}

\def\wde{w_{{\rm de}}}

\def\dl{d_{{\rm L}}}
\def\th{{\rm th}}
\def\obs{{\rm obs}}
\def\diag{{\rm diag}}
\def\de{{\rm de}}
\def\m{{\rm m}}
\def\rhom{\rho_{{\rm m}}}
\def\rhode{\rho_{{\rm de}}}

\def\as{a_{*}}
\def\aap{A$\&$A~}
\def\plb{Phys. Lett. B}
\def\ann{Ann. Appl. Probab.~}
\def\ann{The Annals of Applied Probability}
\def\AJ{Astron. J.~}
\def\mnras{Mon. Non. R. Astorn. Soc.~}
\def\physrep{Phys. Rep.~}
\def\APJS{Astrophys. J. Suppl.~}
\def\APJ{Astrophys. J.~}
\def\prd{Phys. Rev. D~}

\def\PRL{Phys. Rev. Lett.}
\def\JCAP{JCAP}

\usepackage{tabularx}
\usepackage{amssymb,ascmac}
\usepackage{bm}
\usepackage[dvips]{graphicx}

\newcommand{\tbref}[1]{TABLE~\ref{#1}}
\newcommand{\figref}[1]{FIG.~\ref{#1}}
\newcommand{\eqref}[1]{(\ref{#1})}
\begin{document}

\title[Constraining a variable DE model from GRB and SN Ia]{Constraining a variable dark energy model from the redshift-luminosity distance relations of
 gamma-ray bursts and type Ia supernovae}


\author{R. Ichimasa}
\author{E. P. B. A. Thushari}
\author{M. Hashimoto}
\affiliation{Department of Physics, Kyushu University, Hakozaki,
Fukuoka 812-8581, Japan}
\author{R. Nakamura}

\affiliation{Department of Physics, Kurume Institute of Technology, Kamitsu-machi,
Fukuoka 819-0395, Japan}


\begin{abstract}
There are many kinds of models which describe the dynamics of dark energy (DE). Among all
we adopt an equation of state (EoS) which varies as a function of time.
We adopt Markov Chain Monte Carlo method to constrain the five parameters of our models.
As a consequence, we can show the characteristic behavior
of DE during the evolution of the universe.
We constrain the EoS of DE with use of the avairable data of gamma-ray bursts and type Ia supernovae (SNe Ia)
concerning the redshift-luminosity distance relations. 
As a result, we find that  DE is  quintessence-like in the early time and  phantom-like
 in the present epoch or near the future, where the change occurs rather rapidly at $z\sim0.3$.
\end{abstract}


\maketitle

\section{Introduction}\label{int}

Recent observations indicate that our universe is flat and has turned into accelerated expansion phase at the present epoch~\cite{planck14,planck15,riess98,perl99,riess04}.
Some cosmological models have been introduced to examine the characteristics of the present acceleration~\cite{chiba00,cald98,bento02,gao09,hannest,nak08}.
The most simple one is the $\Lambda$CDM model, which includes $\Lambda$ term as a dark energy. 
The $\Lambda$ term leads to the negative pressure, and moderate the universe for acceleration.
This is the standard model of the present cosmology having a dark sector which consists of dark energy and dark matter. 
Dark matter and dark energy should be around 25$\%$ and 70$\%$ at the present epoch, respectively~\cite{planck14}.

Up to now, the $\Lambda$CDM model is almost consistent with many observations of CMB and SN Ia \cite{planck15,suzuki},
 with the exception of the typical estimations of the vacuum energy which are many orders larger than the observed one~\cite{sorkin07}.
Recently, many observational results have been accumulated about SN Ia. Therefore, the compilation data are 
available up to the redshift $z\sim$1.5. 
We have a great interest of investigating the feature of DE around small redshift region. 
While CMB includes the area of very large redshift compared to the one of SNe Ia, 
we may have to interpolate the behavior between them if we study DE quantitatively.
As a consequence, the data $z \geq$1.5 would become important to constrain the behavior of DE in a wide range of cosmological epoch.

Recently, the GRBs have been enthusiastically studied~\cite{yonetoku,amati,capozz,dainotti} to investigate 
 the behavior of DE and the expansion rate at high redshift range. 
As a consequence, we can discuss the density evolution of DE in detail.

Clarifying the properties of DE is one of the most important issues in cosmology, 
and especially modifying an EoS and/or a gravitational field is the most popular method.
Although these are methods to represent the features of DE, it is presumed that DE belongs to dynamical phenomena. 
Some theoretical dark energy models have been proposed to describe the energy density evolution.
For instance, models of quitessence, phantom, quintom, k-essence, chaplygin gas and so on, belong to non-standard DE models~\cite{chiba00,cald98,bento02,gao09,hannest,cai}.
On the other hand, some models which include the modified EoS of dark energy give more direct method. 
We can categorize above models as follows : 
(i) Cosmological constant ($w=-1$). 
(ii) DE with constant but $w \ne -1$. 
(iii) Dynamical DE with $w>-1$ (Quintessence-like models). 
(iv) Dynamical DE with $w<-1$ (Phantom-like models). 
(v) Dynamical DE which acrosses the phantom barrier of $w=-1$ (Crossing models).

Recent observational results indicate that  EoS of DE accrosses the barrier of $w=-1$, so called phantom barrier~\cite{jassal,nesseris,wu,alam}.
Some theoretical models, which accross the phantom divide, have been extensively studied~\cite{koichi,du}.
On the other hand, a modified EoS is easier to handle the density evolution of DE, 
and it is beneficial to understand the asymptotic behaviour of DE to examine whether the crossing exists or not.

In the present work, we investigate how DE should be categorized by modifying the EoS directly.
We adopt a special EoS whose functional form has two limiting values of parameters.
In addition, with use of the observational results such as SN Ia and gamma-ray burst (GRB), we constrain specific parameters in EoS of DE
 over a wide range of the redshift around $1<(z+1)<10$. 
In \S~II, observational data are explained. Our models are presented in \S~III, where mathematical formulation and computational 
method are given. Section IV is devoted to results and discussion.
 

\section{Models}
\subsection{Field equations}
Homogeneous and isotropic universe is described using the Robertson-Walker metric, 
\begin{eqnarray*}
ds^2 = dt^2 - a(t)^2 \left[ \frac{dr^2}{1 - kr^2} + r^2(d\theta^2 + \sin^2\theta d\phi^2  )  \right],
\end{eqnarray*}
where $a(t)$ is the scale factor and $k$ is the curvature constant.
The evolution of our universe is determined from the Einstein equation, 
\begin{eqnarray*}
R_{\mu\nu} - \frac{1}{2}g_{\mu\nu}R = 8\pi G T_{\mu\nu} .
\end{eqnarray*}
Homogeneous and isotropic flow can be regarded as the perfect medium, $T_{\mu\nu}=\diag (\rho,-p,-p,-p)$.
Here, $\rho$ and $p$ are the total energy density and the pressure, respectively.

We can obtain an important equation from the conservation law $\nabla_{\mu} T^{\mu\nu}= 0$ having EoS, $p = w \rho$,
\begin{eqnarray}
\dot{\rho} + 3H(1+w)\rho = 0, \label{c}
\end{eqnarray}
This equation describes density evolution and over-dot indicates ordinary derivative with respect to time. 
From the Einstein equation, we can obtain the following two equations,
\begin{eqnarray}
H^2 &=& \left( \frac{\dot{a}}{a} \right)^2 = \frac{8\pi G}{3}\rho , \label{hubble} \\
\frac{\ddot a}{a} &=& -\frac{4\pi G}{3}(\rho +3p) ,
\end{eqnarray}
where $H$ is  the Hubble parameter, and we take the assumption of flatness ($k=0$).
EoS of each fluid component and/or $w$ should be determined to solve these equations.

We consider that the energy-momentum tensor consists of two fluids ($\rho = \rho_{\de} + \rho_{\m}$): 
(i) $\rho_{\m}$: non-relaivistic matter component as CDM, 
(ii) $\rho_{\de}$: DE with unknown properties.
To study the characteristic feature of DE, we adopt specific EoS of DE, that is, $w_{\de}$ or $w_{de}(a)$
 in EoS which has been proposed by Hannested and M$\ddot{{\rm o}}$rtsell \cite{hannest},
\begin{eqnarray}
\wde (a) &=& \frac{\omega a^\beta + \gamma}{a^\beta + 1},
\end{eqnarray}
where the scale factor $a$ in this EoS is normalized at the time of $\rhom = \rhode$ and $\beta$ is always positive.
Positive (negative) $\beta$ shows the anterograde (retrograde) evolution in terms of the scale factor.
In our choice, $\wde (a)$ converges to $\omega$ at large $a$ and equals to $\gamma$ at the origin.
This EoS can reproduce many kind of DE models mentioned in \S~\ref{int}.

In the present work, we consider the matter and DE as parts of  the energy-momentum tenosor, and these are conserved independently,
\begin{eqnarray}
\dot{\rhode} + 3H (1+\wde)\rhode &=& 0 , \label{c-de} \\
\dot{\rhom} + 3H\rhom &=& 0 , \label{c-m}
\end{eqnarray}
where subscripts 'de' and 'm' indicate  DE and matter, respectively.
With use of EoS and the continuity equations for matter and DE, (7) and (8) can be integrated to obtain
the energy density evolution,
\begin{eqnarray}
\rhom(a)  &=&  \rhom(\as) \left( \frac{a}{\as} \right)^{-3} ,   \\
\rhode(a) &=&  \rhode(\as) \left( \frac{a}{\as} \right) ^{-3} \psi(a), \nonumber \\
    \psi(a;\as) &\equiv& \exp \left( -3 \int_{1}^{a / \as} \frac{\wde (x/\as)}{x}dx \right) = \left( \frac{a}{\as} \right)^{-3\gamma} \left( \frac{a^\beta + 1}{\as^{\beta} + 1} \right)^{-3(\omega - \gamma)/\beta} ,
\end{eqnarray}
where $\as$ is the present value of the scale factor, 
and is not a free parameter but is determined by solving the following equation with a given $\Omega_{\m,0}$,
\begin{eqnarray}
\psi(a=1;\as) = \frac{\Omega_{\m,0}}{1-\Omega_{\m,0}}. 
\end{eqnarray}
Equivalently, $\as$ is the solution of $\Omega_{\m ,0}/(1-\Omega_{\m ,0}) = \as^{3\gamma}[(\as^\beta + 1)/2 ]^{-3(\omega-\gamma)/\beta}$.
We note that $\rhode = \rhom$ at $a=1$.

The density parameters are defined as follows,
\begin{eqnarray}
\Omega_\m(a)  &=& \frac{\rhom(a)}{\rhom(a)+\rhode(a)} ,\\
\Omega_\de(a) &=& \frac{\rhode(a)}{\rhom(a)+\rhode(a)} .
\end{eqnarray}
Indeed, $\Omega_\m = \Omega_\de = 1/2$ at $a=1$.
With the above quantities, the Hubble parameter can be written as follows,
\begin{eqnarray}
&& H(a) = H(\as) \left( \frac{a}{\as} \right)^{-3/2}   \left[ \Omega_{\m}(\as) + \Omega_{\de}(\as)\psi(a;\as)  \right]^{1/2} .
\end{eqnarray}

\subsection{Obsevational data}
SNe Ia are the well known probe of DE due to the measurement of observations of the magnitude-redshift relation up to $z\simeq 1.5$~\cite{riess07,amanullah}.
They are utilized to limit the cosmological parameters. 
In particular, observations for SNe Ia have led to constain the Hubble constant or the density fraction of DE. 
To constrain the cosmological parameters, we adopt the {\it Supernova Union2.1} compilation\cite{suzuki} and {\it Supernova Legacy Survey (SNLS)}\cite{snls} data. 
Moreover, recent analysis of observations indicates that GRBs  can also become the prove of DE. 
Therefore, we employ the redshift-luminosity distance relation obtained from GRB observations 
which are estimated by J. Liu and H. Wei~\cite{liu}.
Cosmological formulas for the luminosity distance $\dl$ and distance moduli $\mu$
(the difference between the apparent and
absolute magnitude) are obtained as a function of the scale factor $a$ as follows,
\begin{eqnarray}
\dl(a) &=& \frac{1}{a}\int_{1}^{a} \frac{dx}{x^2 H(x)}, \\
\mu(a) &=& 5 \log_{10}(\dl (a)/10~ \rm{pc}),
\end{eqnarray}
where $\mu$ is usually shown as a function of the redshift parameter $z$.

\subsection{Computational method of cosmological parameters}
To constrain the present cosmological parameters, following variables are defined,
\begin{eqnarray}\label{14}
&& w_0 = \wde(\as), \;\;
w_a = \left. \frac{d\wde}{da} \right\vert_{a=\as}, \;\;
H_0 = H(\as), \nonumber \\
&& \Omega_{\de,0} = \Omega_\de(\as), \;\;
\Omega_{\m,0} = \Omega_\m(\as),   \;\;
z + 1 = \frac{\as}{a}. 
\end{eqnarray}
Each term having a lower subscript represents the present values respectively for the Hubble parameter,
energy fraction of DE and matter, and the cosmological redshift. 
Here, $(z+1)$ is exactly an unity at present.

We investigate three specific cases for models of dark energy: (i) vEoS: a variable EoS model which has 5 free parameters; $\Omega_{\m,0}$ (equivalently $\Omega_{\de,0}$), $\beta$, $\omega$, $\gamma$ and $H_0$.
(ii) cEoS: constant EoS model in which $\omega$ is always equal to $\gamma$, in this case $\beta$ makes no sence.
(iii) C.C: standard $\Lambda$CDM model which consists of cosmological constant and cold dark matter.

To find the best fit values we calculate $\chi^2$ as follows,
\begin{eqnarray*}
\chi^2 = \sum_{i=({\rm SNe,GRBs})}^{N} \frac{ [\dl {}^{\th}_{,i}(a;\omega,\gamma,\beta,\Omega_{\m,0},H_0) - \dl {}^{\obs}_{,i}(a)]^2 }
                           { \sigma^{2}_{\obs,i}(a)}.
\end{eqnarray*}
Where $N$ is the total number of observational data of SN Ia and GRB.
We can evaluate the best fit values of parameters by minimamizing $\chi^2$ values.

We apply  Markov Chain Monte Carlo (MCMC) method to constrain the parameters of models. 
Here, we define a proposal distribution function $q(\bm{x}' \vert \bm{x})$ which is an arbitrary function,
 and a target distribution $\pi (\bm{x})$. 
The proposal distribution function works better or worse for the convergence steps.

First, we set initial values $\bm{x}^{(0)}=(x^{(0)}_{1},x^{(0)}_2,\cdots,x^{(0)}_N)$ and the step length $\bm{dx}^{(0)}=(dx^{(0)}_{1},dx^{(0)}_2,\cdots,dx^{(0)}_N)$.
Second, we predict the next value $\bm{x}^{(n+1)}$ and the step length $\bm{dx}^{(n+1)}$,
\begin{eqnarray*}
x_{i}^{(n+1)} &=& 
\left\{
\begin{tabular}{ll}
 $x_{i}^{(n)}+ \epsilon_{i}^{(n)}dx_{i}^{(n)}$ & ( $u \le \alpha(\bm{x}^{(n)},\bm{x}^{(n-1)}) $ ) \\
 $x_{i}^{(n)}$ & (otherwise)
\end{tabular}
\right. \\
dx_{i}^{(n+1)} &= &
\left\{
\begin{tabular}{ll}
 $\frac{\vert \bm{dx}^{(n)} \vert \epsilon_{i}^{(n)} dx_{i}^{(n)} }{ \sqrt{\sum_i (\epsilon_{i}^{(n)} dx_{i}^{(n)})^2} } $ & ( $u \le \alpha(\bm{x}^{(n)},\bm{x}^{(n-1)}) $ ) \\
 $dx_i^{(n)}$ & (otherwise)
\end{tabular}
\right.
\end{eqnarray*}
where $u$ is the uniformed random number, $\alpha(\bm{x}^{(n)},\bm{x}^{(n-1)}) = \min\{1, \frac{ \pi(\bm{x}^{(n-1)})q(\bm{x}^{(n)}\vert \bm{x}^{(n-1)})}{\pi(\bm{x}^{(n)})q(\bm{x}^{(n-1)}\vert \bm{x}^{(n)})}\}$ is the acceptance probability, 
 $\pi(\bm{x}^{(n)})$ and $\pi(\bm{x}^{(n-1)})$ are the $\exp(-\chi^2/2)$ values of each step.
$\bm{\epsilon}$ are random numbers according to the normal distribution. 
In the present work, we assume $q(\bm{x}\vert\bm{x}^{(n-1)})= q(\bm{x}^{(n-1)}\vert\bm{x})$ 
beacuse of the assumption of detailed balance, and $\alpha(\bm{x}^{(n)},\bm{x}^{(n-1)})$ is reduced to a simple formula: 
$\alpha(\bm{x}^{(n)},\bm{x}^{(n-1)})= \min\{1,\pi(\bm{x}^{(n-1)})/\pi(\bm{x}^{(n)})\}$.
We preserve the data point of $x^{(n)}$ for the case of $u \le \alpha(\bm{x}^{(n)},\bm{x}^{(n-1)}) $, 
 and the next predicted value is $x^{(n+1)}_{i} = x^{(n)}_{i} + dx_i $.

Note that we set the norm of $\bm{dx}$ as a constant value in $N$-dimensional parameter space and 
 $\bm{dx}$ is redefined when $\bm{x}^{(n+1)} \ne \bm{x}^{(n)} $. 
Moreover, the step length $\epsilon_{i}^{(n)}dx_{i}^{(n)}$ are wighted by normal distribution function. 

We set 100 bins from the minimum to the maximum value for each $x_i$, and we performed MCMC calculation till $\pi({\bm x})$ are converged.

\section{Results and discussions}
\subsection{Best fit parameters}
As the first evaluation, we search the best fit parameters for each model. We show the best fit parameters and $\chi^2$ values for each model in TABLE~\ref{bestfit}.
Particularly, in vEoS model, we have found that the parameter $\beta$ takes the range as $\beta > 20$, and $\omega$ prefers less than $-1$ and $\gamma$ prefers greater than $-1$.
This indicates that the feature of DE should be changed drastically. 
Both $H_0$ and $\Omega_{\m,0}$ seem to be consistent with the Planck 2015 results.
Furthermore, the energy density of DE is only slightly affected from these parameters of the early stage in the universe.
Since DE cannot become the candidate to solve the cosmological constant problem, the value of the (effective) EoS may be increased by some unknown mechanisms in the earlier epoch.
%
\begin{table}[h]
\begin{center}
\caption{Best fit parameters for three models and corresponding $\chi^2$ values.}
\label{bestfit}
\begin{tabular*}{0.8\hsize}{@{\extracolsep{\fill}}cccc}
\hline
\hline
Parameter           & vEoS & cEoS & C.C. \\
\hline
$\omega$                 & -1.02                  & -1.03  & -1  \\
$\gamma$                 & -0.873                 & -1.03  & -1  \\
$\beta$                  & $> 20$                 & --     & --  \\
$H_0$                    & 70.1                   & 69.9   & 69.8  \\
$\Omega_{\m,0}$          & 0.2801                 & 0.2993 & 0.2897  \\
\hline
$w_0$                    & -1.02                  & -1.03  & -1  \\
$w_a$                    & -1.73$\times 10^{-6}$  & 0      & 0  \\
\hline
$\chi^2_{\rm min}(\Delta \chi^2)$  & 726.9                  & 728.9(+2.0)  & 729.4(+2.5)  \\
\hline
\end{tabular*}\\
$\beta$ is the arbitrary value and $\omega=\gamma$ for the models of cEoS and C.C.
\end{center}
\end{table}

\begin{figure}[h]
\begin{center}
\begin{tabular}{c}
  \begin{minipage}{0.3\hsize}
    \includegraphics[width=\hsize]{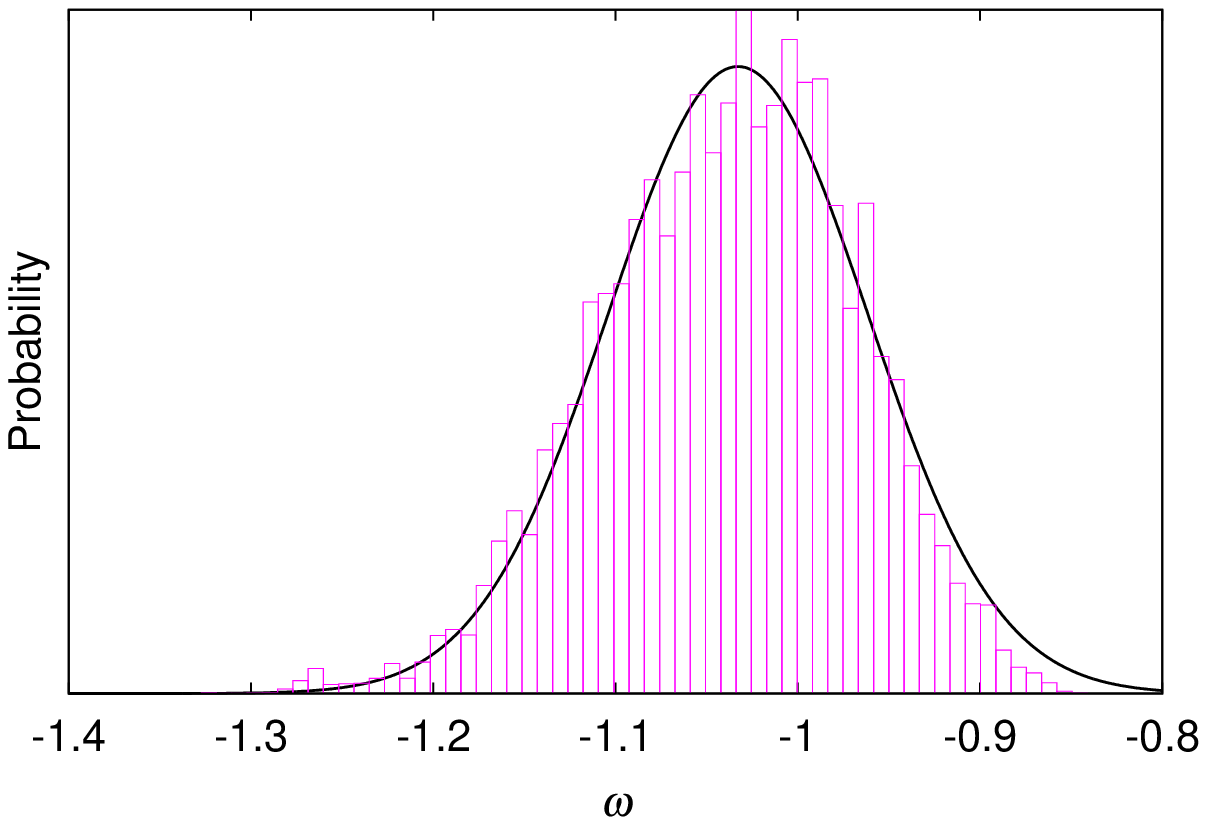}
  \end{minipage}
  \begin{minipage}{0.3\hsize}
    \includegraphics[width=\hsize]{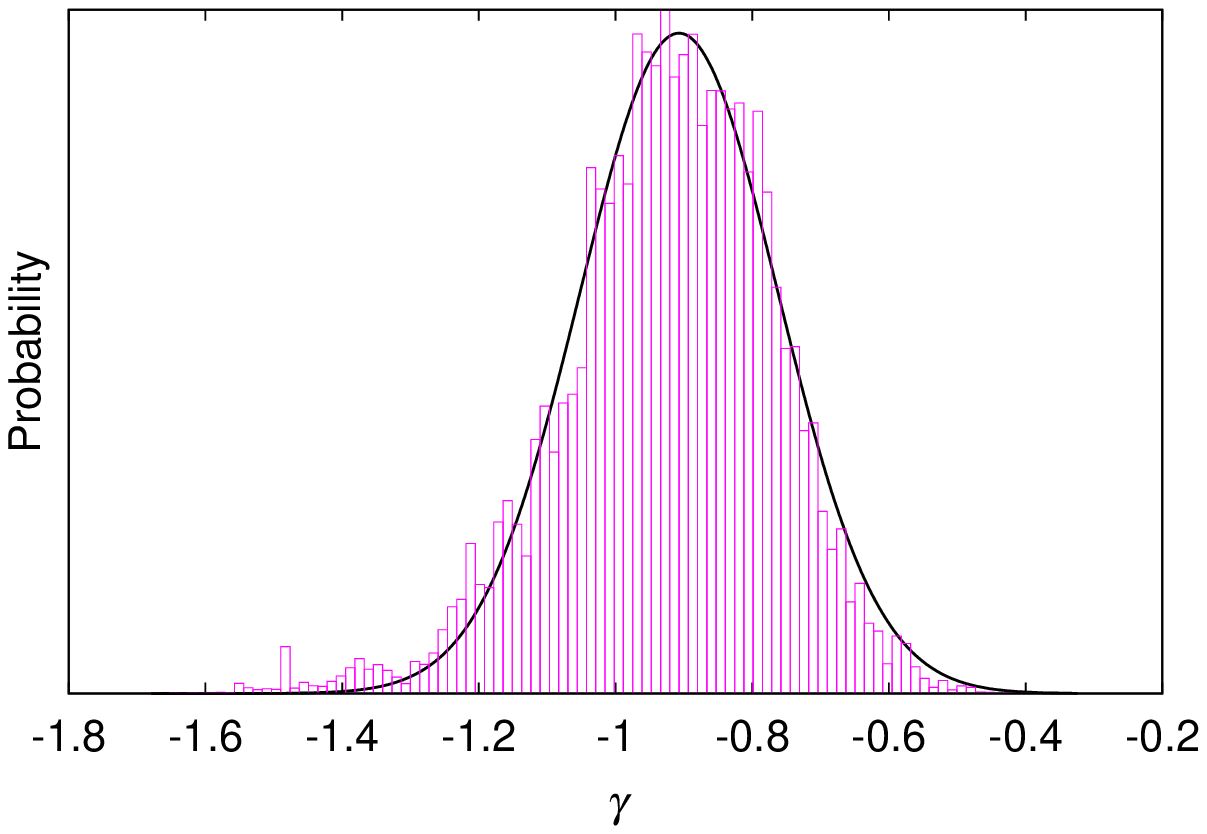}
  \end{minipage}
\\
  \begin{minipage}{0.3\hsize}
    \includegraphics[width=\hsize]{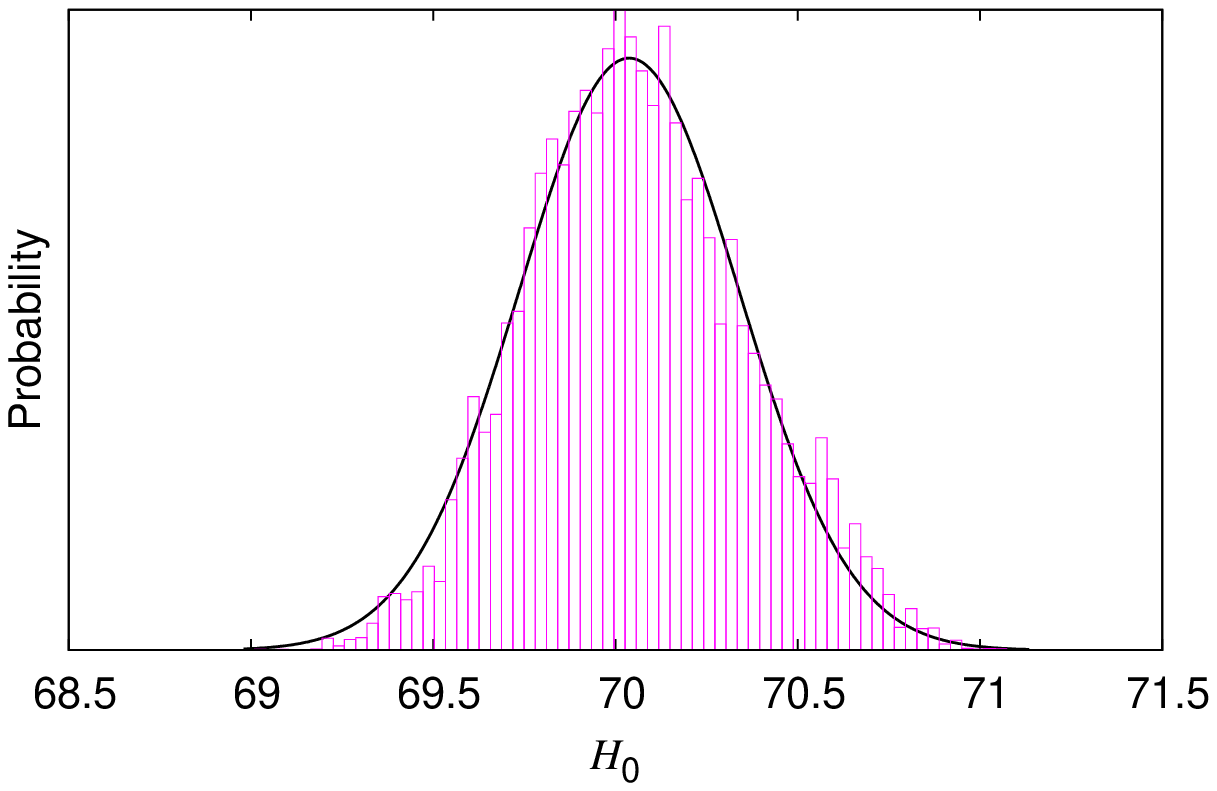}
  \end{minipage}
  \begin{minipage}{0.3\hsize}
    \includegraphics[width=\hsize]{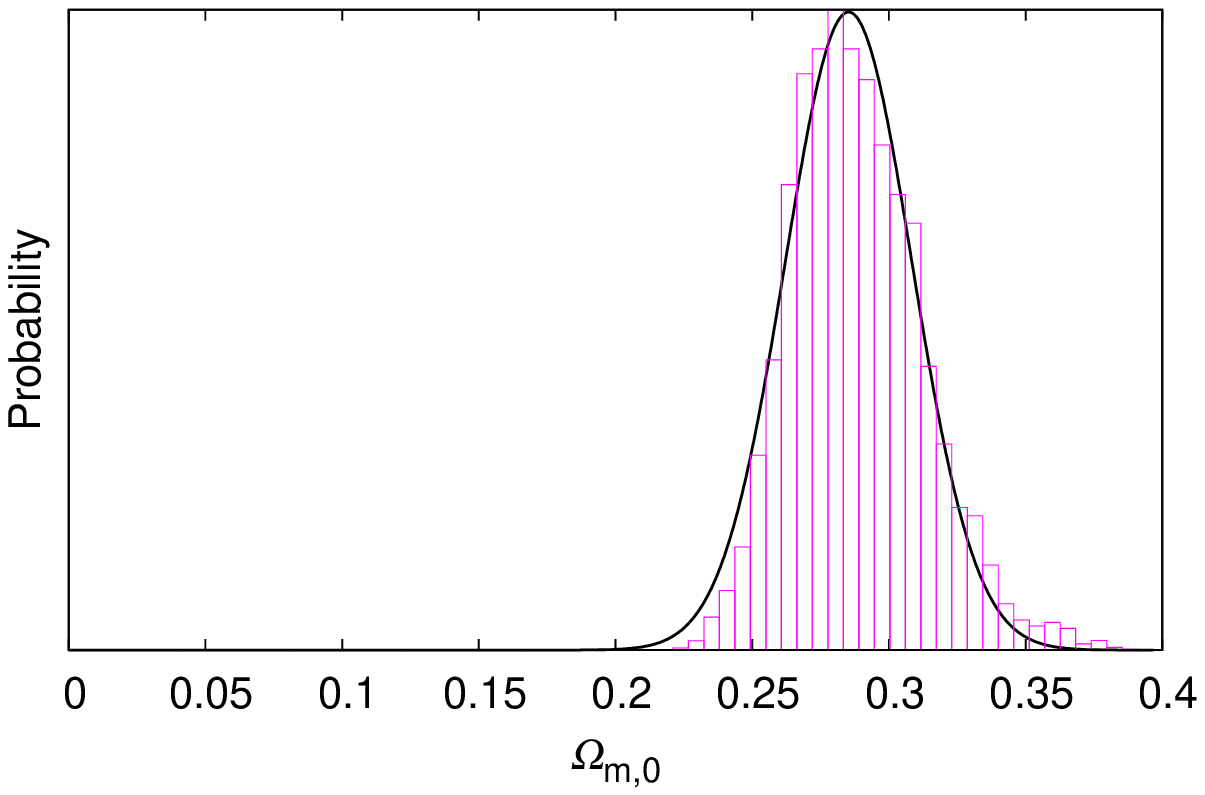}
  \end{minipage}
\end{tabular}\\

\end{center}
\caption{Posterior distributions for $\omega$, $\gamma$, $H_0$ and $\Omega_{m,0}$. 
The boxes and lines show data points and the plausible functions as Gaussian, respectively.
These constraints come from the SNe Ia and GRBs observations. }
\label{dist}
\end{figure}

\begin{figure}[h]
\begin{center}
\includegraphics[width=0.5\linewidth]{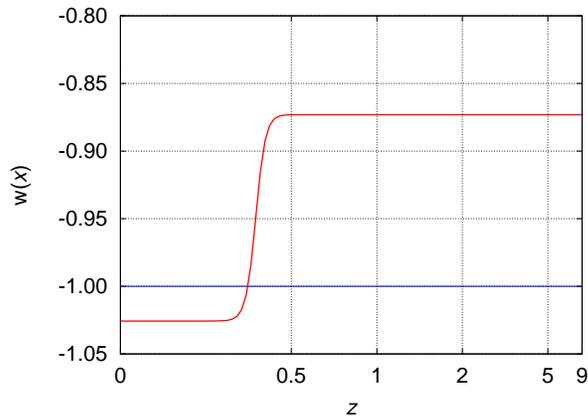}
\end{center}
\vspace{0.5cm}
\caption{EoS of DE as a function of time. Holizontal line (blue) shows cosmological constant 
and curved line (purple) shows variable EoS with best fit parameters. We find that DE changes its property from 
quintessence-like in the early time  
to phantom-like field in the present epoch at $z \sim 0.3$ or equivalently $a \sim 0.75$.}
\label{eos}
\end{figure}

For the next step, we apply MCMC method to obtain a reliable reagion for each parameter as shown in \figref{dist}.
This result indicates that DE should change the property from quintessence-like to phantom-like field at $z \sim 0.3$ (see \figref{eos}).
From the result that $w_a \sim 0$ and $\beta$ is very large, it would be more important
to search the turning point ($z \sim 0.3$) than to evaluate the slope of EoS at present.
In fact, some models whose (effective) EoS gives the similar condition are constructed in terms of the modified gravitational theory (e.g., \cite{bam09,cai07,du}).

Let us discuss the convergency of the MCMC method.
Since the acceptance rate is affected by not only the proposal distributions $q(\bm{x} \vert \bm{x}^{(n-1)})$ but also dispersion parameters (or step lengths),
 dispersion parameters for each Gaussian should be estimated in burn-in period. 
Theoretically, it is known that an ideal acceptance rate of the random walk algorithm in $N$-dimensions is about $23.4\%$~\cite{rob97}.
However, our method need not the evaluation of dispersions but we have to determine only $\vert \bm{dx}^{(0)} \vert$, 
 and step lengths are redefined automatically depending on the previous accepted step of $\epsilon_{i}dx_{i}$. 
In the present study acceptance rate was $\sim 25\%$.
Therefore, our method with the parameter settings has operated rather well.

\subsection{Evaluation of the models}
\begin{figure}[h]
\begin{center}
\includegraphics[angle=270,width=0.5\linewidth]{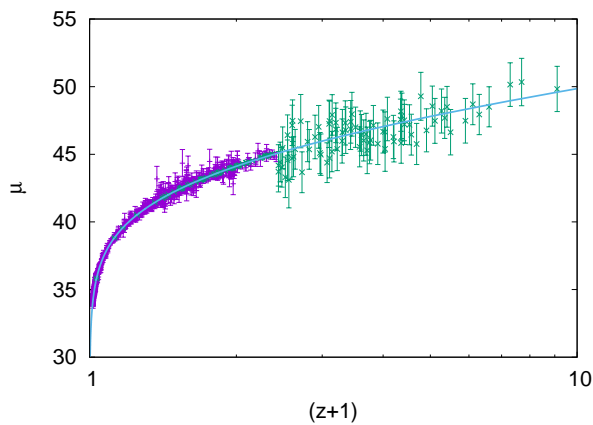}
\end{center}
\caption{Magnitude-redshift relation for the vEoS model. The observational data are given with error bars;
purple: SNe Ia~\cite{suzuki,snls}, green: GRBs~\cite{liu}.}
\label{m-z}
\end{figure}

We have investigated the properties of DE by analysing the observational data sets of SNe Ia and GRBs
as shown in \figref{m-z}, where we note that the difference due to parameters given in \tbref{bestfit} is very small even for
$z\sim 10$.

In order to compare models in \tbref{bestfit} with the standard model (C.C.), we adopt Akaike information criteria (AIC)\cite{aic}. 
AIC is applied to the models with a defferent number of free parameters. 
We will define AIC such as ${\rm AIC} = \chi^2_{\rm min}+2n$, where 
$n$ is the number of free parameters. 
Here, we define the difference between C.C. and the other model $i$ as $\Delta ({\rm AIC})_{i} = ({\rm AIC})_{i} - ({\rm AIC})_{{\rm C.C.}}$. 
The negative value of $\Delta ({\rm AIC})_{i}$ proves the priority of the model $i$ compared to the C.C.
In previous studies, SNe Ia data indicate that $\wde$ may cross the $-1$ barrier.
However, we find that $\Delta ({\rm AIC})_{\rm vEoS}=3.5$ and $\Delta ({\rm AIC})_{{\rm cEoS}}=1.5$, 
and therefore we cannot have clear evidence (see Table 7 and 8) for the time dependency of DE.

\begin{figure}[ht]
\begin{center}
\includegraphics[width=0.5\linewidth]{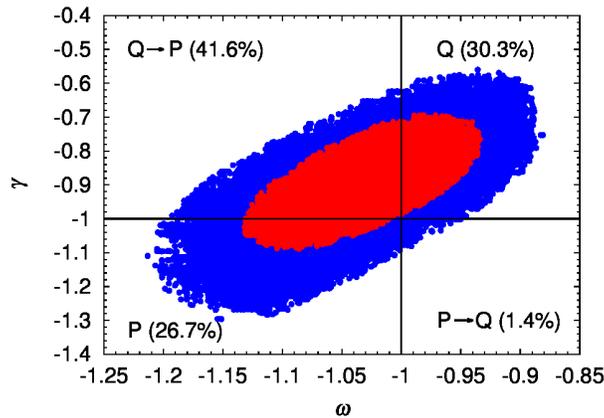}
\end{center}
\vspace{0.5cm}
\caption{Constraints obtained from probability distribution on $\omega$ and $\gamma$;
 68$\%$ C.L. and 95$\%$ C.L. correspond to red and blue regions, respectively. 
The area  on a plane of $\gamma$ against $\omega$ is divided into four square regions.
Q and P denote quintessence-like and phantom-like models. The crossing whose DE evolves from Q to P
is indicated by Q$\rightarrow$P. P$\rightarrow$Q shows that the crossing proceeds oppositely.}
\label{w-g}
\end{figure}

\figref{w-g} shows the probability distribution on the plane of $(\omega,\gamma)$.
We define 68$\%$ C.L. and 95$\%$ C.L. such as $\Delta \chi^2 = 2.30$ and $\Delta \chi^2 = 6.18$, respectively.
The area  on a plane of $\gamma$ against $\omega$ is divided into four square regions.
Q and P denote quintessence-like and phantom-like models. The crossing whose DE evolves from Q to P
is indicated by Q$\rightarrow$P. P$\rightarrow$Q shows that the crossing proceeds oppositely.
If the parameter set $(\omega,\gamma)$ exists in the upper right region, DE always behaves quintessence.
The lower left region belongs to phantom.
The percentages denote the integrated probability in each separated (squared) region.
This indicate that the model whose DE evolves from $w_{\de}<-1$ to $w_{\de}>-1$ (P$\rightarrow$Q) is excluded in 2.4 $\sigma$ confidence level.

On the other hand, we can conclude that the transition from $w_{\de}>-1$ to $w_{\de}<-1$ would occur rapidly around $z \sim 0.3$ if the crossing exists (see \figref{eos}).
Some crossing models have been already constructed (e.g., \cite{bam09,cai07,du}), and the EoS changes from $w_{\de}>-1$ to $w_{\de}<-1$ moderately in these models.
Our results may indicate that the alternation would occur more instantly, because $\beta$ should be taken a large value.

Finally, we obtain the following results with $68\%$ C.L. : 
$\omega = -1.03 \pm 0.11$, $\gamma = -0.91 \pm 0.14$, $H_0 = 70.0  \pm 0.3  $ and $\Omega_{m,0} = 0.285 \pm 0.023$.
$H_0$, $\Omega_m$, $\Omega_\Lambda$, and $w_{\de}(a)$ are consistent with the Planck 2015 results (XIII) of the  $\Lambda$CDM model~\cite{planck15}(TT,TE,EE+lowP+lensing).
On the contrary, the Planck result (XIV, Fig. 5) ~\cite{planck15xiv} assuming EoS to be represented a first-order of Taylor expansion of $w(z)$, 
 EoS evolves from $w_{\de}<-1$ to $w_{\de}>-1$. The redshift dependency of the EoS is completely opposite direction compared with our result.

Other investigations~\cite{suzuki,hannest} indicate that EoS evolves from $w_{\de}>-1$ to $w_{\de}<-1$, which
are the same tendency compared with our result.
Hannestad et al.~\cite{hannest} adopt CMBFAST package. They insist that it is hard to constrain more than two parameters from 157 "gold" samples of SN Ia data only. 
Therefore, they utilize SN Ia + LSS (SDSS and 2 degree Field Galaxy Survey;2dFGRS) + CMB (WMAP) data.
However, latest 695 SN Ia data~\cite{suzuki,snls} can constrain five parameters, and gives sumaller value $(\chi^2_{\min}/{\rm d.o.f.} = 0.97)$ than 
 that by Hannestad et al. ($\chi^2_{\min}/{\rm d.o.f.}=1.10$; SNI-a best fit model in Table 2~\cite{hannest}).
Combining 695 SN Ia and 138 GRBs data analysis result in a smaller value $\chi^2/{\rm d.o.f.}=0.87$. 
In conclusion, we succeed in constraining five parameters with no presumption of parameter range. 

Our results of $w_a$ in Eq.~\eqref{14} are consistent with those obtained from
Union 2.1 (see Table 7 and 8 in ~\cite{suzuki}) whose data is limitted to SNe Ia.
Our studies with use of only SNIa data~\cite{suzuki} give $w_{\de} \simeq -1.02$ for $z<0.3$ and $w_{\de} \simeq -0.97$ for $z\sim 0.5$ with $\Omega_m =0.277$.
Using the values $w_0$ and $w_a$ given in \cite{suzuki} ($w_z$CDM and SNe+CMB in Table 7), the EoS value is around $w \sim -0.96$ at $z=0.5$ with $\Omega_m = 0.273$ from Eq.~(7) in Ref.~\cite{suzuki}.

In the present work, an equality epoch concerning matter and dark energy 
is changed by $\delta z \sim +1.5$ compared to the case of C.C. due to inclusion of SNIa and GRB
observations, which affects the formation of the first object.

\begin{acknowledgements}
We thank Dr. Kenzo Arai for helpful discussion.
This work has been supported in part by a Grant-in-Aid for Scientific Research (24540278, 15K05083) of the Ministry of Education, Culture, Sports, Science and Technology of Japan.
\end{acknowledgements}


\end{document}